\documentclass[
superscriptaddress,
twocolumn,
amsmath,amssymb,
aps,
prb,
]{revtex4-2}

\usepackage{graphicx} 
\usepackage{dcolumn} 
\usepackage{bm} 
\usepackage{braket}
\usepackage{color}
\usepackage{amsthm}
\usepackage{hyperref} 

\definecolor{lightblue}{RGB}{73,151,208}
\definecolor{crimson}{RGB}{140,41,53}

\hypersetup{
    colorlinks,
    linkcolor={crimson},
    citecolor={lightblue},
    urlcolor={lightblue}
}

\newtheorem{definition}{Definition}[section]

\newtheorem{theorem}{Theorem}[section]

\begin{document}

\preprint{APS/123-QED}

\title{Quantum advantage for noisy channel discrimination} 

\author{Zane M. Rossi}
\affiliation{%
Department of Physics, Massachusetts Institute of Technology, Cambridge, Massachusetts 02139, USA}%
\author{Jeffery Yu}
\affiliation{%
Department of Physics, Massachusetts Institute of Technology, Cambridge, Massachusetts 02139, USA}%
\author{Isaac L. Chuang}
\affiliation{
Department of Physics, Department of Electrical Engineering and Computer Science, and Co-Design Center for Quantum Advantage,
Massachusetts Institute of Technology, Cambridge, Massachusetts 02139, USA}
\author{Sho Sugiura}
\affiliation{%
Physics and Informatics Laboratory, NTT Research, Inc.,
940 Stewart Dr., Sunnyvale, California, 94085, USA 
}%

\date{\today}%

\begin{abstract}
    Many quantum mechanical experiments can be viewed as multi-round interactive protocols between known quantum circuits and an unknown quantum process.
    Fully quantum ``coherent'' access to the unknown process is known to provide an advantage in many discrimination tasks compared to when only incoherent access is permitted, but it is unclear if this advantage persists when the process is noisy.
    Here, we show that a quantum advantage can be maintained when distinguishing between two noisy single qubit rotation channels.
    Numerical and analytical calculations reveal a distinct transition between optimal performance by fully coherent and fully incoherent protocols as a function of noise strength.
    Moreover, the size of the region of coherent quantum advantage shrinks inverse polynomially in the number of channel uses, and in an intermediate regime an improved strategy is a hybrid of fully-coherent and fully-incoherent subroutines.
    The fully coherent protocol is based on quantum signal processing, suggesting a generalizable algorithmic framework for the study of quantum advantage in the presence of realistic noise.
\end{abstract}

\maketitle

\section{Introduction} \label{sec:introduction}
    
    Experimental progress over the past twenty years has increasingly enabled the coherent manipulation of complex quantum mechanical systems, bolstering the ongoing search for settings where quantum protocols permit advantage over their classical counterparts. This progress has both informed and been informed by the development of novel quantum algorithms. For many such algorithms it is assumed that a multiplicity of unitary operations can be coherently applied to a prepared quantum state, and indeed numerous results support the intuition that quantum advantage often relies on the ability to perform deep quantum circuits.
    
    Multiple recently developed frameworks have considered quantum advantage based on quantum and classical access models \cite{qualm_2021, Huang21}. In particular, for certain inference problems, it has been shown that a quantum advantage is permitted to models wherein a quantum process can be applied coherently, versus models without such coherent access. This approach, first fixing a problem and then comparing algorithmic performance across \emph{differing access models}, has permitted novel complexity-theoretic insights into sources of quantum advantage.
    
    A missing piece in much of the work on access model dependent quantum advantage is an explicit and constructive study of the effect of noise. In \cite{qualm_2021}, the exponential query complexity advantage studied does not survive the introduction of noise, while relatedly the quantum advantage studied in \cite{Huang21} is proven for finite noise, but only information theoretically. In an era of noisy quantum devices, it will be important to understand the gap between these two approaches: constructive investigations of the effect of noise for realistic inference tasks, and a fuller understanding of which problems permit a performance gap among access models to persist under finite noise.
    
    This work aims to understand this gap between \cite{qualm_2021} and \cite{Huang21} by investigating an instance of quantum advantage which (1) is constructive, (2) incorporates finite noise, and (3) employs specific parameters which enable a map of where this advantage exists (and does not), given reasonable assumptions. 
    Specifically, while \cite{Huang21} demonstrates an information theoretic no-go theorem for quantum advantage in average case regression tasks for machine learning, we choose a classification problem, to which their results are not directly applicable.
    Moreover, while \cite{qualm_2021} considers comparisons between noiseless settings and unparameterized noise, we incorporate noise which is parameterized by a single, continuous value.
    With our binary classification task and simply parameterized noise, we are able to identify, and visually and compellingly depict, regions of quantum advantage with respect to signal and noise parameters, illustrating novel thresholds.
    
    The specific problem we consider is discrimination among two noisy single qubit rotation channels, where the noise is defined by classical distributions over the rotation angle. Given consistent access to one among two possible quantum channels, where sampling rotates the querent's qubit by said noisy angle, the querent is challenged to determine which distribution underlies their sampling power. 
    This problem is perhaps the simplest instance of a more general class of quantum channel discrimination problems, which are known to be difficult and rely on sophisticated use of quantum resources (e.g., entanglement, auxiliary space).
    The normally distributed noise we consider is entirely parameterized by its mean and standard deviation, and consequently plotting the the performance of coherent and incoherent access protocols against these two parameters reveals the thresholds alluded to above.
    
    We find that up to a certain noise threshold, a coherent access protocol can always outperform its incoherent access counterpart for our hypothesis testing problem, but that the reverse is true above this threshold. Moreover, below the threshold, we show that there exist a family of even better performing hybrid protocols, which are alternately coherent and incoherent. That is, for such protocols we find that one should compute coherently for a certain time, measure, and repeat, and we compute the optimal query complexity, or \emph{coherence length}, for the coherent subroutines of these hybrid protocols.
    
    The problem proposed in this work, in addition to relating to reasonable noise models, is one for which the recently developed methods of quantum signal processing (QSP) \cite{low-16, low-19, low-chuang,gilyen} are natural. Indeed, much of our analysis relies on the application of known properties and guarantees of QSP, though we here extend these methods to a new, noisy context. 
    
    This work is structured as follows: in Section~\ref{sec:noiseless_discrimination} we discuss a noiseless instance of the hypothesis testing problem introduced in Definition~\ref{def:rdg}, for which optimal quantum protocols are known, and generalize these results to the case of noise in Section~\ref{sec:noisy_discrimination}, where various limits permit closed form analysis of the behavior of these discrimination protocols. Finally, we examine hybrid protocols in Section~\ref{sec:large_n_limit} before discussing their significance.

        \subsection{Problem statement} \label{subsection:problem_statement}
            
    		We formalize the problem statement sketched in the previous section, fixing an instance of quantum inference; specifically, we consider symmetric hypothesis testing among quantum channels. Let two distinct distributions be $\Theta_0, \Theta_1$, over the reals. These distributions have well-behaved probability density functions denoted $\Theta_b(\theta)$ for $b \in \{0, 1\}$. We involve these distributions in an ``RDG'' game involving quantum channels, as given in Definition \ref{def:rdg}.
    		
    		\begin{definition} \label{def:rdg}
    		    Rotation discrimination game (RDG). A party with a single qubit is afforded oracle access to a single qubit quantum channel taking the following form: when queried, some $\theta$ unknown to the party is drawn from $\Theta_b$  (either 0 or 1, with $b$ fixed for all queries), and the unitary channel
                    \begin{equation}
                        \mathcal{E}_b \equiv \exp{(i\theta \sigma_x)}
                    \end{equation}
    		   is applied to the party's qubit, where $\sigma_x$ is the Pauli $X$ operator with determinant $1$. 
    		    
    		   The distributions $\Theta_b$ are taken to be normal distribution \footnote{One could instead consider the wrapped version of these distributions, as the map to quantum channels is periodic.}, defined by two parameters 
    		   \begin{equation} \label{eq:normal_dist}
    				\Theta_b(\theta) 
    				= \frac{1}{\sqrt{2\pi \sigma^2}} e^{-\frac{(\theta_b - \mu_b)^2}{ \sigma^2}},
    			\end{equation}
    		    where $\mu_b$ and $\sigma$ are the mean and the standard deviation respectively. The challenge is to, in as few queries as possible, determine which classical distribution parameterizes the quantum channel; i.e., to determine $b$ with high confidence and low query complexity. The total query complexity is denoted $N$. We use a shorthand notation for the separation of the means of the distributions, $\delta = |\mu_0 - \mu_1|$.
        		
        		Moreover, the party is assumed to only apply \emph{serial, non-adaptive protocols}, i.e., they have access to only one qubit, and their strategy must be independent of intermediate measurement results. The resources otherwise afforded to an algorithm playing an RDG, specifically \emph{coherent access} or \emph{incoherent access}, are defined below.
        	\end{definition}
    
            This problem is simple to describe, yet sufficiently rich to exhibit a distinct transition between optimal performance of coherent and incoherent access models.
            
            More specifically, we give this problem because it will be natural to consider two limits in the single noise parameter $\sigma$ and the mean separation $\delta$. One limit is the near noiseless case, i.e., $\sigma \rightarrow 0$; in the absence of noise, the coherent model is strictly better than the incoherent model, which we discuss in Section~\ref{sec:noiseless_discrimination}. The other limit is when the noise is much larger than the mean angular separation $\delta$, or simply when $\delta \rightarrow 0$. In this limit coherence is almost immediately lost, and a fully incoherent protocol will show advantage because it can measure more often. As the protocols we consider will encompass well-performing strategies for both of the above limits, they will be useful in characterizing the difficult-to-analyze intermediate regions.
    		
\begin{figure}[htpb]
        		    \centering
        	        \includegraphics[width=0.65\columnwidth]{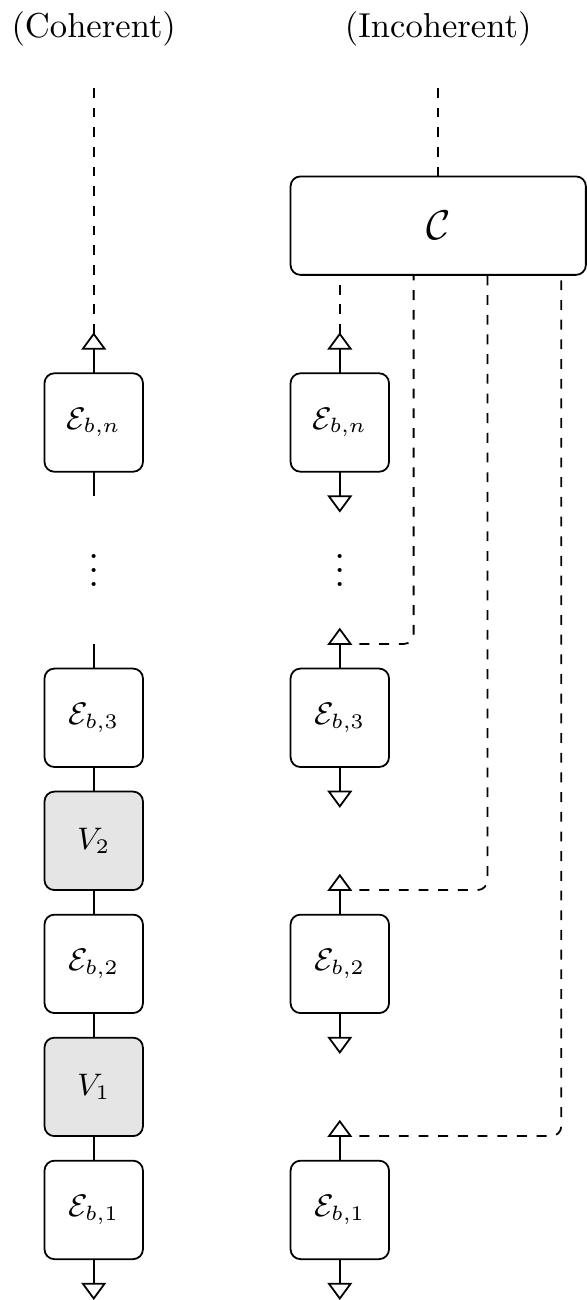}
        		    \caption{Non-adaptive serial incoherent access (right) and coherent access (left) protocols for RDGs. Here $\mathcal{C}$ is some classical thresholding procedure performed on the results of measurements, dotted lines indicate transmission of classical information, and triangles are complete measurements when leaving a channel application and classically controlled quantum state preparations going into a channel application. The $V_j$ are unitary quantum gates, while $\mathcal{E}_{b,j}$ is the $j$-th application of $b$-th hypothesis channel. This figure follows conventions established in \cite{qualm_2021}, in which time progresses as one moves up the diagram.}
        		    \label{fig:access_models}
        		\end{figure}

    		As stated before, we are interested in performance among quantum algorithms situated in coherent and incoherent access models, for specific tasks. We want to make concrete the distinction between quantum strategies for RDGs in these two models, depicted in Figure \ref{fig:access_models}. Before this, however, we give a short definition. In the single qubit setting a \emph{complete measurement} will mean simply a rank one POVM, defined by $\{\lvert \psi\rangle\langle \psi\rvert, I - \lvert \psi\rangle\langle \psi\rvert\}$ for some pure single qubit state $\lvert \psi \rangle$. Equivalently a complete measurement is one that prepares quantum states which are completely defined by measurement results. We now define our access models of interest.
    		
    			\begin{itemize}
    
    				\item \textbf{Incoherent access:} The querying party is forced to perform a \emph{complete measurement} between each channel application. Measurement outcomes are processed through some classical thresholding procedure.
    
    				\item \textbf{Coherent access:} The querying party may now defer complete measurement, and may instead perform intermediate quantum gates on the qubit  between successive applications of $\mathcal{E}_b$.
    
    			\end{itemize}
    	
    		The protocols we consider will, as stated in the RDG definition, take place in a serial, non-adaptive setting regardless of access model. That is, the single qubit channel $\mathcal{E}_b$ is not applied jointly to many qubits, and quantum operations are performed independent of intermediate measurement results. This is a restriction, but one which does not remove all interesting properties of the comparison.
    		
    	\subsection{Prior work} \label{subsection:prior_work}
    	    
    	    Much of the work discussing quantum advantage in relation to coherence does so for problems in inference \cite{helstrom, acin, Huang21, Zhou21}. Statistical inference, a central and broad tool in experimental contexts, has been investigated with respect to quantum mechanical systems for over fifty years, and hypothesis testing among quantum channels specifically, the subject of this work, has been studied in a variety of contexts for the past twenty years \cite{duan, duan_feng_ying, pirandola, zhuang_pirandola}. While binary hypothesis testing in quantum settings has been characterized for both quantum states \cite{helstrom} and restricted sets of quantum channels \cite{acin}, such initial treatments did not consider general noisy settings. Moreover, these seminal results are still being elaborated today to pertain to greater contexts, most often the presence of noise and adaptiveness.
    	    
    	    The specific approach of analyzing quantum algorithmic performance according to access model is more recent \cite{qualm_2021}. Along with the complexity-theoretic treatment of \cite{qualm_2021}, a similar distinction has been investigated for machine learning problems \cite{Huang21}, differentiating between the performance of coherent access and incoherent access protocols for worst-case approximation of specific observables. In the latter work, as their inference task involves all $n$-qubit Pauli operators, which can be efficiently measured in the coherent access model by clever constructions, a performance separation was proven via information theoretic methods against any, even adaptive incoherent access protocol. Surprisingly, the work of \cite{Huang21} guarantees the persistence of this quantum advantage even in the presence of noise of any strength. While their problem is one of regression, inferring expectation values averaged over noise, in contrast a new challenge comes about when one tries perform classification. Such problems, primarily investigated in \cite{qualm_2021}, can lead to a vanishing of quantum advantage under finite noise. It is the fragility of quantum advantage for classification problems, and its apparent robustness in regression problems, that presents an interesting gap. Expanding such robustness to classification remains an intriguing obstacle to characterize and overcome.
    	    
    	    Given that quantum mechanics is a resource convex theory \cite{takagi_general, takagi}, there exist discrimination problems for which any additional resource can provide algorithmic advantage. However, investigating algorithmic advantage among access models, even for simple access models and tasks, is in general a difficult problem, and often approached by non-constructive methods. Therefore, providing constructive evidence for any gap in performance, even within relatively restrictive access models can provide insight into the benefits of coherence. Moreover, if these restrictions can be made while maintaining physical motivation, then conclusions can serve as a good basis for future, still constructive generalizations.
    	
    	    Various bounds already exist for the performance of algorithms for quantum hypothesis testing among noisy quantum channels, though these results rely on novel information-theoretic proofs in complicated resource models \cite{zhuang_pirandola, pirandola}. While it is surprising and intriguing that these results exist, most still do not consider relative performance among (1) coherent and incoherent access models, and (2) for classification problems specifically. For even simple classification games like the RDGs we consider, no existing work investigates the relative advantage between coherent and incoherent access models.
    	    
    	    In a seemingly entirely separate subfield of quantum algorithms, the development of quantum signal processing (QSP) \cite{low-16, low-19, low-chuang} and the quantum singular value transform (QSVT) \cite{gilyen} has encouraged a fresh look at the commonalities underlying notable quantum algorithms \cite{mrtc_2021_unification}. These algorithmic families, which permit the efficient polynomial transformation of eigenvalues of unitary operators and singular values of embedded linear operators respectively, have proven flexible: they are able to reproduce optimal quantum algorithms for problems as diverse as Hamiltonian simulation, the quantum linear systems problem, and factoring \cite{gilyen, mrtc_2021_unification}. Since the polynomial transformations within QSP and QSVT can be efficiently performed even on unknown unitary processes, they are expected to be powerful methods for addressing inference problems, though application to noisy settings has been limited thus far. Building this bridge between quantum algorithms and quantum inference in noisy settings (e.g. from QSP to RDGs) is thus of interest, both for understanding constructive methods to achieve quantum advantage for inference problems, as well as for developing new scenarios for quantum algorithms.

\section{Noiseless discrimination} \label{sec:noiseless_discrimination}

    Before discussing solutions to the major problem discussed in this work, i.e., the discrimination of noisy quantum processes in the form of RDGs (Definition \ref{def:rdg}), it is worthwhile to discuss the noiseless case. This simpler problem is not only amenable to closed form results but also provides intuition that will help guide us through the introduction of noise.

    It is the aim of this section to answer the following questions in the noiseless case, as each will become important and non-trivial in the general setting.
        
        \begin{itemize}
            
            \item Do there exist pairs of quantum channels for which coherent access (e.g., QSP-based) discrimination methods provably outperform incoherent access protocols, and what is the quantitative nature of this relative speedup?

            \item What is the simplest resource model in which an optimal discrimination protocol is possible with respect to choice of channel hypotheses?

        \end{itemize}
        
    We consider these two questions below, first analytically, then with a concrete example.
        
    \subsection{Sufficiency of QSP protocols for optimal noiseless discrimination}    
        
    To investigate these questions we state a simple channel discrimination problem concretely in the noiseless setting, employing Definition \ref{def:noiseless_rdg} as a noiseless RDG setting.
        
        \begin{definition} \label{def:noiseless_rdg}
            \emph{Noiseless RDGs}. We consider one concrete instantiation of an RDG (Definition \ref{def:rdg}): distinguishing between two quantum channels $\mathcal{E}_0, \mathcal{E}_1$ with explicit form $\exp\{i\theta_0\sigma_x\}$ and $\exp\{i\theta_1\sigma_x\}$ respectively, where $\sigma_x$ is a Pauli operator as before, and each of $\theta_0, \theta_1$ is fixed.

            A party is given the ability to apply a quantum channel $\mathcal{E}_b$ for some consistent $b \in \{0, 1\}$ without knowledge of $b$. The party is again tasked with the following goal: determine, given repeated access to $\mathcal{E}_b$, the hidden bit $b$. 
        \end{definition}
        
	In reference to general RDGs, in this game the distributions $\Theta_b$ over angles result in unitary channels $\mathcal{E}_b$, and are thus Dirac distributions peaked $\mu_b = \theta_b$. In other words, this is the problem of discriminating two fixed known rotations. The problem of unitary channel discrimination, for which this problem is one example, has been studied for both coherent and incoherent access protocols \cite{helstrom, acin}. This section translates these results into the language of QSP protocols, and provides a new statement for the separation in performance of coherent access and incoherent access protocols, toward analysis of the more complicated case where each $\mathcal{E}_b$ is non-unitary (Section~\ref{sec:noisy_discrimination}).
	
	It turns out that in this setting we have a complete characterization of both coherent access and incoherent access protocols (in the non-adaptive case). To make this more clear, we present a new family of quantum protocols.
	
	    \begin{definition} \label{def:qsp_strategies}
			A \emph{QSP protocol} for an RDG is defined by (1) a series of QSP phase angle lists, $\{\Phi_1, \Phi_2, \cdots, \Phi_m\}$, each of which is in $\mathbb{R}^{r_j}$ for $j \in \{1, 2, \cdots, m\}$ and $r_j$ a positive integer, and (2) a series of classical descriptions of preparations and projective measurements $\{(\psi_1, \psi_1^\prime), (\psi_2, \psi_2^\prime), \cdots, (\psi_m, \psi_m^\prime)\}$.

            Here the probability to measure $\rvert \psi_{j} \rangle$ is $\lvert \langle \psi_j^\prime \lvert Q_{\Phi_j} \rvert \psi_{j} \rangle \rvert^2$, and  $Q_{\Phi_j}$ has the form
				\begin{eqnarray}
					Q_{\Phi_j} = e^{i\phi_0\sigma_z}\prod_{\ell = 1}^{r_j}\left(\mathcal{E}_b\,e^{i\phi_\ell\sigma_z}\right).
				\end{eqnarray}
            The total number of $z$-rotations is equal to the number of channel applications, i.e., $\sum_j r_j = N$.
		\end{definition}

    Evidently QSP protocols have non-trivial intersection with coherent access protocols, and moreover, when $m = 1$, they are a proper subset of coherent access protocols. It is also not difficult to see that in the case of $r_j = 1$ the QSP protocol given in Definition \ref{def:qsp_strategies} reduces to an incoherent access protocol. For the incoherent access protocol we will consider in comparison with QSP protocols, $N$ classical measurement outcomes are processed through a majority vote. Regardless, when comparing coherent and incoherent access protocols, we will optimize the QSP angles, preparations and projective measurements such that the error probability is minimized. QSP protocols are an important and sufficient subset of quantum algorithms for RDGs to discuss, as will be shown.

    It is instructive to cast our procedure in the framework of \cite{qualm_2021}, and contrast our results. The problem they consider concerns the inference of an unknown quantum channel from among two possibilities. Their framework also includes noisy inference problems, but the advantage they prove is for a noiseless variant. Their quantum circuit for a coherent access protocol, like ours, comprises a set of possible gates, while in the incoherent case any two unknown channel applications must be interrupted by local operations and classical communication (LOCC). The majority vote we introduce can always be implemented with LOCC.  Moreover, QSP protocols also fit within the framework of \cite{qualm_2021}.  Therefore our discrimination procedure fits within their framework.  On the other hand, our setting is simpler, and includes noise.  We show quantum advantage for the noiseless case in this section, and in Section~\ref{sec:noisy_discrimination} we go further than \cite{qualm_2021}, by showing a quantum advantage which persists in the presence of finite noise.
		
	Now back to QSP protocols for RDG.  Given well-known results for the form of QSP protocols, the ease of analysis of QSP sequences, and the flexibility of polynomials constructed using QSP, they provide an excellent starting point for the analysis of quantum hypothesis testing in the presence of noise. Using QSP, we can state the following theorem for noiseless RDGs.
    	
    	\begin{theorem} \label{theorem:noiseless_qsp}
    		For noiseless RDGs (Definition \ref{def:noiseless_rdg}), coherent access protocols can always match or exceed the performance of incoherent access protocols. Moreover, there exists a finite positive integer $N$ and a coherent access non-adaptive protocol using $N$ queries such that this protocol perfectly decides the bit $b$ naming the hidden channel, where $N = \mathcal{O}(\delta^{-1})$ is optimal.
    
            \begin{proof}
                The existence of such an $N$ follows from the results of \cite{acin} under the recognition that this is a unitary channel discrimination problem, and direct construction can be found in \cite{rossi2021quantum}. That coherent access protocols can always outperform incoherent access ones follows from the latter strictly containing the protocols comprising the former. Thus we have shown at least the sufficiency of QSP protocols for optimal noiseless discrimination.
            \end{proof}
    	\end{theorem}

    It is worthwhile to explain why the fully coherent QSP protocol given in Theorem \ref{theorem:noiseless_qsp} performs obviously better than its incoherent access counterpart. We motivate a simple and old result from quantum information which will appear again and again in analyzing the basic behavior of incoherent access protocols.
            
    The one-shot distinguishability of two unitary quantum channels $\mathcal{E}_0, \mathcal{E}_1$ is determined by the maximum over initial density operators $\sigma$ and all POVMs, and gives the minimum error probability
        \begin{equation}
            p_{err} = \min_{\sigma}\frac{1}{2}\left(1 - \frac{\lVert \mathcal{E}_0(\sigma) - \mathcal{E}_1(\sigma)\rVert}{2}\right),
        \end{equation}
    where $\lVert\, \cdot\, \rVert$ is the trace distance, the implicit optimal POVM the \emph{Helstrom measurement}, and the overall statement the \emph{Helstrom bound}. For noiseless RDGs, this bound takes the form
        \begin{equation}
            p_{err} = \frac{1}{2}(1 - \sin \,\lvert \theta_0 - \theta_1 \rvert),
        \end{equation}
    where it will be taken without loss of generality that $\delta = \lvert \theta_0 - \theta_1\rvert \leq \pi/2$. This simple bound completely defines the performance of incoherent protocols up to polynomial factors when paired with post-processing on measurement results, and will serve to constitute one performance bound in the comparison to coherent access protocols.
    
    One possible post-processing method in the incoherent access setting is the majority vote; this work will solely be considering this type of non-linear post-processing, though many additional classical statistical methods and thresholding procedures are possible, though none of them qualitatively changes the statements made for advantage in the rest of the work.
    
    We define this majority vote operation below and discuss its performance in comparison to QSP protocols for noiseless RDGs. A majority vote (often denoted MAJ) unsurprisingly returns the majority result from a set of $(2M + 1)$ i.i.d.\ Bernoulli samples defined by some underlying success probability $1/2 \leq p \leq 1$. The distribution which defines the output of a majority vote is itself a Bernoulli distribution with a modified $p$. I.e., it corresponds to a single sample returned from the modified Bernoulli distribution defined by
        \begin{equation}
            p^\prime = \sum_{k = 0}^{M} \binom{2M + 1}{k} p^{2M + 1 - k}(1 - p)^{k},
            \label{eq:maj-vote}
        \end{equation}
    which for the constraints given satisfies $p^\prime \geq p$ when $M > 0$, an integer. An analogous reversed statement can be made for $0 \leq p < 1/2$.
        
    For incoherent access protocols we consider that the querent is forced to completely measure after each use of the quantum channel, but is however free to choose this measurement as well as the preparation of the input to which the channel is applied. In this restricted setting the Helstrom bound can be combined with majority vote to give a more complete picture of the performance of incoherent access protocols.
    
    For coherent access protocols, in comparison, we consider instead that the querent may successively apply the quantum channel multiple times to a chosen input, possibly interspersing these applications with quantum operations of their own choosing before finally measuring with respect to a chosen complete measurement. While it is clear that the use of the hypothesis channel in a QSP sequence is a subset of such protocols, it is not clear that optimal discrimination protocols can be performed serially and without entanglement. However, this is demonstrated to be the case in Theorem \ref{theorem:noiseless_qsp}, as well as for many noiseless channel discrimination protocols \cite{duan_feng_ying}, and this paper concerns itself with the extension of such a reasonable notion of optimality to reasonably noisy settings.
    
    Together with the statement that coherent protocols can always outperform incoherent ones in the noiseless setting, we are thus interested in an upper bound for the magnitude of the relative advantage between the two protocols with respect to a particular pair of quantum channels. One such bound is discussed in a concrete, simple example below.
    
    \subsection{A simple concrete example}
        
        \begin{figure}
            \includegraphics[width=\columnwidth]{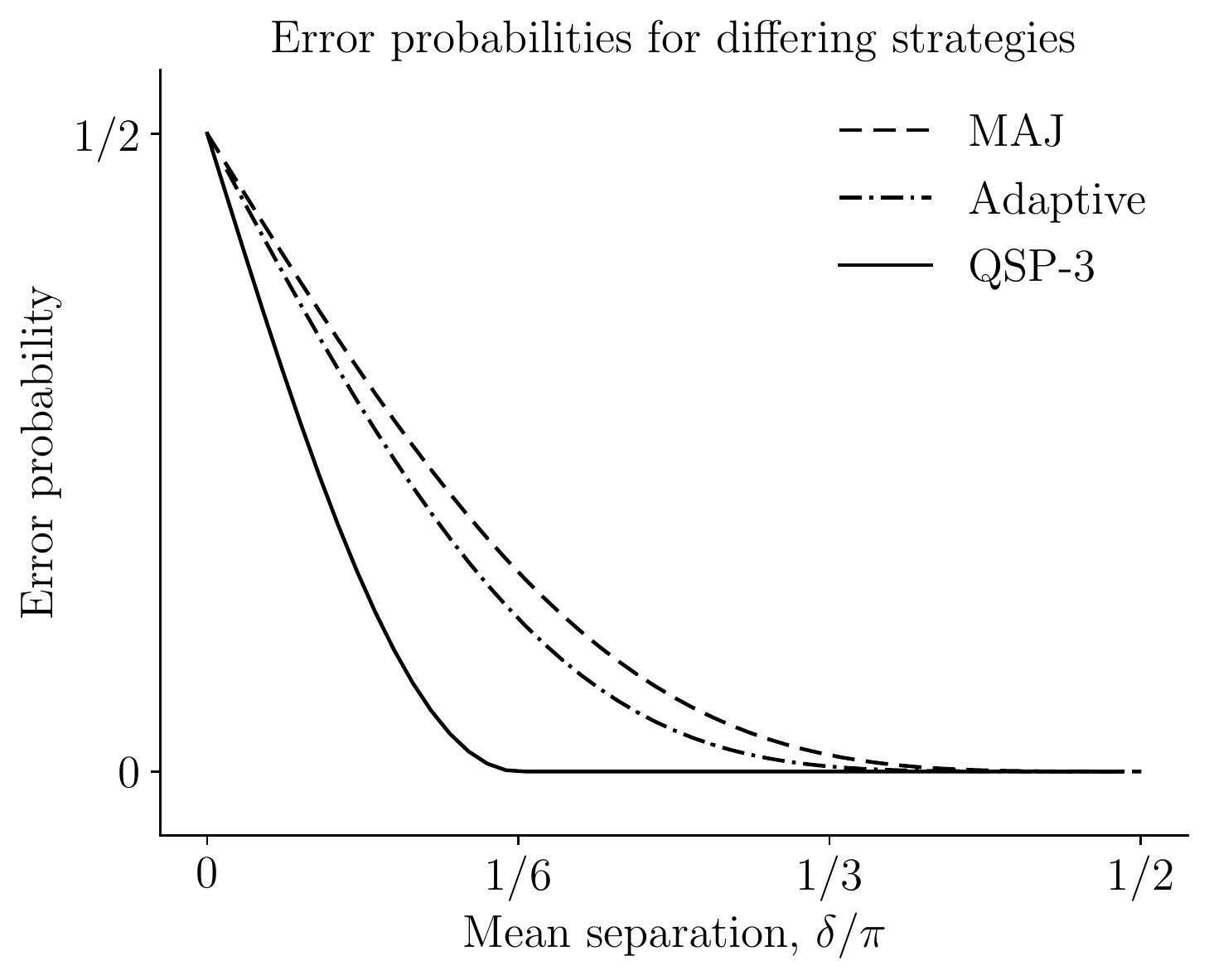}
            \caption{Error probability is plotted as a function of distance between signal means $\delta$. We compare protocols which use three total queries ($N = 3$), and which are either fully coherent (solid), fully incoherent and non-adaptive (dashed), and fully incoherent and adaptive (dot-dashed).}
            \label{fig:qvh0_theory}
        \end{figure} 
        
        If we consider a concrete problem and employ constructive QSP protocols for its solution, the corresponding performance can be explicitly analyzed. In Figure \ref{fig:qvh0_theory} we compare the performance of two methods for symmetric binary hypothesis testing for rotations about a fixed axis in the $N = 3$ case. The same functional form derived here will extend easily to larger $N$.
        
        Considering the performance of (possibly suboptimal, but only polynomially so) incoherent protocols first, the probability of error from successive Helstrom measurements followed by majority votes can be calculated explicitly, taking $N = 2M + 1$ an odd positive integer
            \begin{equation} \label{eq:repeated_helstrom}
                \begin{split}
                \text{MAJ}(p_{\rm err})
                =\,
                &2^{-(2M + 1)}\sum_{k = 0}^{M} \binom{2M + 1}{k}\\
                &\times\left(1 - \sin\delta\right)^{2M + 1 - k}
                \left(1 + \sin\delta\right)^{k}
                \end{split}
            \end{equation}
        for $\delta \in [0, \pi/2]$, where $\text{MAJ}$ is the majority vote function on $p_{\rm err}$, the one shot probability of incorrectly distinguishing the two channels with the Helstrom measurement.
        
        We can compare the the behavior of (\ref{eq:repeated_helstrom}) with the corresponding behavior of an optimized QSP sequence, which for this noiseless setting is provably optimal in query complexity, and has the simple form, for a fixed $N$,
            \begin{equation} \label{eq:simple_qsp}
                p_{\rm err}
                =
                \frac{1}{2}(1 - \sin{N\delta})
            \end{equation}
        for $\delta = \lvert \theta_0 - \theta_1 \rvert \leq \pi/N$ and zero otherwise (in which case the results of \cite{acin, rossi2021quantum} enable perfect discrimination). It is not difficult to show that this function is always strictly less than (\ref{eq:repeated_helstrom}) for any positive choice of $N$ and any non-zero separation $\delta$. The magnitude of the ratio of the error probabilities is discussed in Figure \ref{fig:qvh0_theory}, and can for certain choices of parameters be arbitrarily large.
        
        Although we do not discuss adaptive protocols in the rest of this work, we do depict the performance an adaptive strategy for an incoherent access protocol in Figure ~\ref{fig:qvh0_theory}. The procedure is as follows: After each measurement, the posterior probability of hypothesis $\Theta_b$ is updated by Baye's theorem. Given these updated probabilities, one can numerically optimize the projective measurement used the next run so that mutual information between the probability of $\Theta_b$ and measurement's defining probability is maximized. The details of this method are explained in Appendix~\ref{appendix:adaptive_strategy}. The adaptive incoherent protocol is strictly better than plain majority vote, but still cannot achieve perfect discrimination when the hypothesis channels are not orthogonal.
    
\section{Noisy discrimination}\label{sec:noisy_discrimination}
    
    The consideration of noise in channel discrimination is essential, given that realistic quantum computers exhibit noise, and more interestingly that the introduction of noise can impart important nuance on statements of advantage for specific quantum algorithms \cite{qualm_2021, robust_noise_ibm}.
    
    This section explores if and when the observed gap in performance between coherent and incoherent access protocols for RDGs discussed in the previous section is robust to finite noise. We will show that even for simple, parameterized noise, this question leads to a transition boundary for quantum advantage among access models.  We present these findings in two subsections below, first numerically, then analytically.
        
    \subsection{Quantum advantage for noisy discrimination}
        
        \begin{figure*}
            \centering
               \includegraphics[width=\textwidth]{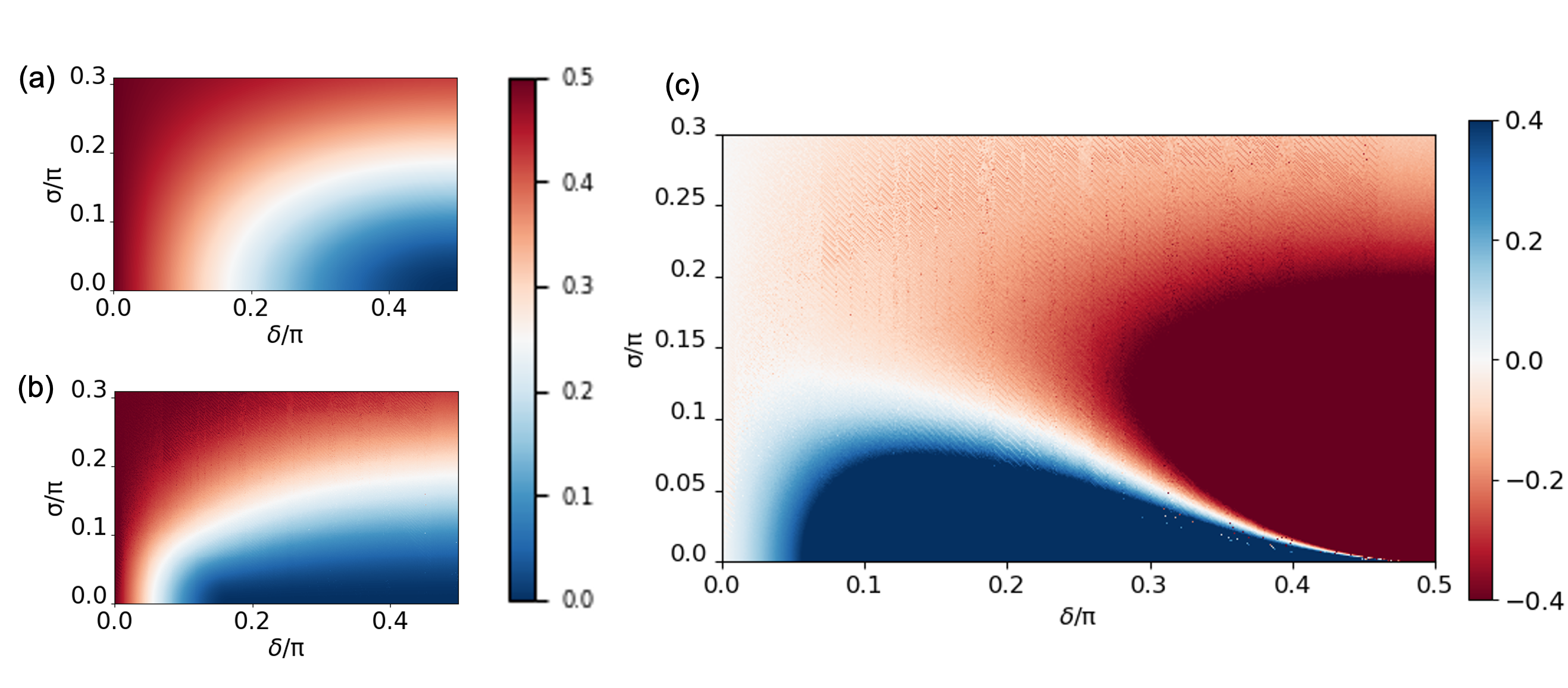}
             \caption{(a, b) Error probabilities for MAJ-3 (a) and QSP-3 (b). In (c) is depicted the ratio of the error probability of MAJ-3 and that of QSP-3. Red is used for the area where  MAJ-3 performs better, white indicates equal performance, and blue indicates where QSP-3 performs better. Quantum advantage for coherent access protocols is present when both the `signal' ($\delta$) and the `noise' ($\sigma$) are not too large.}
            \label{fig:noisy_case}
        \end{figure*}  
        
        In QSP, we can tailor the sequence of phase angles which define the QSP sequence to optimize for discriminating between two channels parameterized by classical distributions, $\Theta_0$ and $\Theta_1$. When the channels are noiseless, i.e., $\sigma=0$,  QSP protocols have already been shown to have better success probabilities than incoherent protocols for discrimination. 
        
        In the limit of large noise, we might expect that a coherent protocol can do no better than its incoherent counterpart. Consequently an interesting regime is when, by some reasonable metric, the signal to noise ratio for the quantum channel is roughly unity. We investigate this region, and give evidence for a sharp transition boundary for relative quantum advantage in the parameter space defining the underlying noise. Additionally, we analytically determine this boundary in two distinct limits where its computation is somewhat simplified.
        
        Where we cannot analytically determine this transition boundary, for $RDG_{\delta, \sigma}$ with $\sigma > 0$, we use Monte Carlo methods to numerically optimize over QSP angles as described in Appendix~\ref{appendix:numerical_methods}. The result for length three QSP protocols (QSP-3) is shown in Figure~\ref{fig:noisy_case}(b). For a fair comparison we need to fix the number of channel applications (the query complexity) afforded to each quantum protocol, so we compare with MAJ-3 in Figure~\ref{fig:noisy_case}(a). In a side-by-side comparison, it is evident that the QSP-3 protocol has a lower error probability in regions of low $\delta$ and low $\sigma$. On the other hand, however, both protocols perform poorly in regions of high noise, though this is expected since in the limit of uniform $\Theta_0, \Theta_1$, the distributions become indistinguishable. 
        
        To visualize the region of the quantum advantage more clearly, we show the ratio of the success probabilities of QSP-3 versus MAJ-3 in terms of $\delta$ and $\sigma$ in Fig~\ref{fig:noisy_case}. The darker blue region indicates the region of advantage, and later analytic results will show that this region is convex, and extends to finite $\delta$ and $\sigma$.
    
    \subsection{Transition boundaries in the low-separation limit}\label{sec:transition_low_separation}

        In the previous subsection the relative performance of QSP and incoherent access protocols was investigated numerically. To cross-check the validity of these results, as well as extend them to a region (i.e., low separation, or low $\delta$) in which the simulations are numerically unstable, we can consider the limit of the multi-dimensional integrals which define these success probabilities.

        For a noisy RDG, when the anglular separation between the means of the possible rotations is sufficiently small, the optimal phase angles of a QSP protocol tend to zero (i.e., toward the \emph{simple QSP protocol} which simply rotates along a fixed great circle on the Bloch sphere, and performs the Helstrom measurement). We give a definition below.
        
        \begin{definition}
            Simple QSP protocol. A simple QSP protocol is one which uses the trivial choice of $\Phi = \{0,0,\, \cdots, 0\}$ as well as a projective measurement which optimizes discrimination success for a particular pair of hypothesis channels.
        \end{definition}
        
        In this case, we can analytically calculate the expected success probability even in the noisy case and thus identify the transition boundary between coherent and incoherent access protocols. We first compare coherent and incoherent access protocol performance for a fixed odd query complexity $N = 2M + 1$. In this setting it is not too difficult to compute the expected error probability of a simple QSP protocol
            \begin{equation} \label{eq:qsp_strategy_performance}
                \overline{p_{\rm QSP}}
                =
                \frac{1}{2}\left(1 - (2M + 1)\delta  e^{-2(2M + 1)\sigma^2}\right),
            \end{equation}
        where the angular separation is small enough that $\sin{\delta} \approx \delta$. Relatedly, we can take the $M = 0$ instance of (\ref{eq:simple_qsp}), the Helstrom bound, and apply the majority vote  (Eq.~\eqref{eq:maj-vote}), again keeping only leading order terms in $\delta$,
            \begin{equation} \label{eq:majority_vote_performance}
                \overline{p_{\rm MAJ}}
                =
                \frac{1}{2}\left(1 - \frac{[2M + 1]!}{[M!]^2}  2^{-2M} \delta e^{-2\sigma^2} \right),
            \end{equation}
        which is computed by summing the relevant binomial terms in the expansion of the majority vote function of the Helstrom success probability (see Appendix \ref{appendix:maj_vote_proof}). Comparing these two in the limit of small angular separation $\delta \ll \pi/2$, one finds the condition for the transition boundary in terms of $\sigma$,
            \begin{equation} \label{eq:transition_boundary}
                \sigma = \frac{1}{2}\sqrt{\frac{1}{M}\ln\left(\frac{2^{2M}(M!)^2}{(2M)!}\right)}, 
            \end{equation}
        which can be shown by approximation to scale as $\sigma^2 \approx \log{(M)}/M = \log{(N)}/N$ as $N$ grows sufficiently large. For $M = 1$ (the length $3$ sequences discussed at length in this paper's numerical results), this boundary occurs at $\sigma = \sqrt{\ln{2}}/2$, which agrees well with numerical simulation. Qualitatively, we see that the region of advantage, for a fixed noise, grows smaller with increasing query complexity; one might naturally expect this as the Bloch sphere is compact, and thus a standard application of the law of large numbers cannot be applied to arbitrarily accurately threshold about a mean.
    
\section{Hybrid protocols for large N and optimal coherence length} \label{sec:large_n_limit}
    
    We now pose a separate, related, and natural question regarding the protocols for RDG which we have been investigating. We will show that this question leads to yet another limit under which the transition region discussed above becomes analytically tractable and instructive.
    
    Imagine that the querying party has some large budget for the total number of queries they can make to the quantum process defining some RDG. It is known from the previous sections that, if one knows that the underlying distributions for the two possible quantum channels are relatively narrow, then a coherent access protocol is best, while for sufficiently large noise an incoherent access protocol does well. 
    If the querying party has knowledge of the magnitude of the channel noise then how should they choose to split the difference between coherent and incoherent access protocols? How long should they compute before measuring, where this length (in terms of query complexity) will be termed the \emph{coherence length}. To make this question concrete we need one more definition.
    
        \begin{definition} \label{def:hybrid_strategies}
            Hybrid protocols. A $\xi$-hybrid protocol for an RDG is one which, given a total budget of $N$ queries to one among two quantum channels, performs $\xi$-length QSP protocols a total of $N/\xi$ times, followed by a majority vote on the $N/\xi$ measurement outcomes. While stating nothing of the methods for finding optimal $\xi$-hybrid protocols, it is easy to see that coherence lengths $\xi = N$ and $\xi = 1$ correspond to coherent and incoherent access protocols respectively.
        \end{definition}
    
    If we consider again (1) the limit of small separation between the means of the distributions defining the two possible quantum channels (i.e., where simple QSP protocols are optimal), and (2) the large $N$ limit, then the work we did before can be re-purposed for differing $\xi$.
    
    I.e., if we suitably scale (\ref{eq:qsp_strategy_performance}) and (\ref{eq:majority_vote_performance}) by $\xi$, we recover that in this careful limit, as $N \rightarrow \infty$, that the error probability for a $\xi$-hybrid protocol goes as
        \begin{equation}
                \frac{1}{2}
                \left(
                1 - \xi\delta\frac{[N/\xi]!}{[(N+1)/2\xi)!]^2}  2^{-(N + 1)/\xi} e^{-2\xi\sigma^2} 
                \right)
            \label{eq: p hybrid}
        \end{equation}
    where $N$ has been taken to replace $2M + 1$ in (\ref{eq:majority_vote_performance}) for simplicity, and is the total number of queries. This can easily be minimized over $\xi$ (and in fact this minimum is unique), where $\xi_{\rm min}$ satisfies the relation
        \begin{equation} \label{eq:optimal_xi}
            -4 \sigma^2 \xi_{\rm min}^2 + N \left(\psi\left[\frac{N + \xi_{\rm min}}{2\xi_{\rm min}}\right] - \psi\left[\frac{N}{2\xi_{\rm min}}\right]\right) = 0
        \end{equation}
    where $\psi$ is the digamma function. Note that we have assumed that the limit of large $N$ has permitted suitable analytic continuation of the discrete objects in (\ref{eq:majority_vote_performance}). We now hope to further explain what solutions for $\xi_{\rm min}$ look like in various reasonable limits.
    
    For large noise, and consequently small coherence length, we can suitable expand the digamma functions in (\ref{eq:optimal_xi}) about $1$, in which case they act logarithmically, i.e., $\psi(1 + x) \approx x$, leading to the solution
        \begin{equation} \label{eq:large_sigma_xi}
            \xi_{\rm min} \approx \frac{1}{4}\sigma^{-2},
        \end{equation}
    which tracks with intuition; in the limit of large noise (a classical limit) the optimal coherence length decreases according to the inverse variance. For the small noise limit, the approximate solution for (\ref{eq:optimal_xi}) becomes a bit more involved, but we can expand the expression given in (\ref{eq:optimal_xi}) about $\xi = N$, in which case we find that for small deviations about $\sigma = 0$, that the optimal coherence length instead goes as
        \begin{equation} \label{eq:small_sigma_xi}
            \xi_{\rm min} \approx \frac{1}{2}\sigma^{-1}\sqrt{ N\left(\frac{\pi^2}{6} + \log{4}\right)}.
        \end{equation}
    We see that it has a dependence on $N$, as we might expect, and a qualitatively different scaling in $\sigma$. That said, in this limit the optimal coherence length can often exceed $N$, in which case a fully coherent protocol is simply optimal.
    
\begin{figure}[htpb]
	    \centering
        \includegraphics[width=\columnwidth]{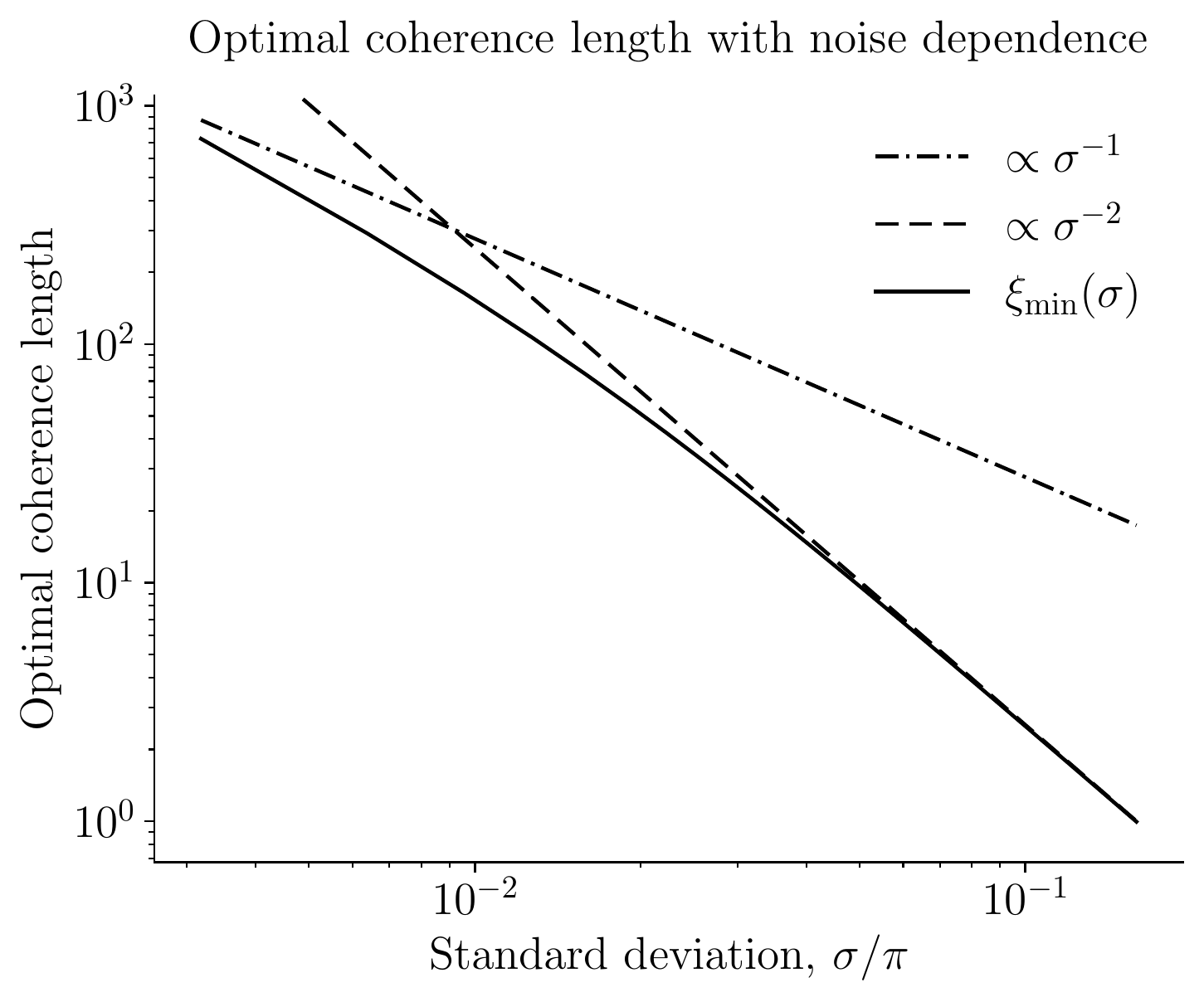}
	    \caption{Optimal coherence length for a hybrid simple QSP sequence for increasing noise, using a log-log scale. Note that only in the limit of large noise does the coherence length fall as one would classically expect, $\xi \propto \sigma^{-2}$. For small noise the slope implies a $\xi \propto \sigma^{-1}$ relation. These limits are derived for (\ref{eq:large_sigma_xi}) and (\ref{eq:small_sigma_xi}) respectively. This plot shows the case for $N = 100$ as an explicit example. Note that the analytically continued `optimal coherence length' $\xi_{\rm min}$ can go above $100$ and below $1$.}
	    \label{fig:optimal_coherence_length}
	\end{figure}

    This transition between polynomial scaling behavior of the optimal coherence length, depicted in Figure \ref{fig:optimal_coherence_length}, is surprising; while the inverse square behavior in a region of high noise tracks with our intuitive understanding of classical statistics, for small noise the slower decay of the optimal coherence length seems to bolster the notion that, at least in this region, the problem displays a more quantum behavior.
    
    In Figure \ref{fig:optimal_coherence_length}, numerical results for the optimal coherent length (solid), its closed form expression in the large $\sigma$ limit (dotted line), and the same in the small $\sigma$ (dotted dashed line) limit are shown. The dependence $\xi \propto \sigma^{-2}$ in the large $\sigma$ region indicates that the optimal length is, as mentioned previously, determined classically, as the coherence length is such so that the accumulated error is kept constant. However, the optimal length always decreases slower than this classical limit, approaching $\xi \propto \sigma^{-1}$ for small $\sigma$. This demonstrates that the discrimination power of a hybrid protocol is augmented by coherent access. 
    
    Note finally that \eqref{eq:optimal_xi} was obtained by analytic continuation of the factorial function, and thus, $\xi_{\rm min}$ is not bounded. In Fig.~\ref{fig:optimal_coherence_length}, $\xi_{\rm min}$ becomes smaller than one around $\sigma/\pi = 0.2$, indicating that an incoherent access protocol gives a minimum error probability. Similarly, a fully coherent protocol becomes optimal around $\sigma/\pi = 0.01$. Evidently a hybrid protocol is optimal between these limits. 

\section{Conclusion and Discussion} \label{sec:discussion}
        
    In this work we show a quantitative difference in performance amongst quantum protocols for hypothesis testing against pairs of noisy quantum channels. The two families of quantum protocols considered for this task, coherent and incoherent access protocols, were constructively instantiated as QSP protocols (Definition \ref{def:qsp_strategies}) and incoherent access non-adaptive protocols. For these specific protocols, in the case of rotation discrimination games (RDGs; Definition \ref{def:rdg}), we provided analytic arguments for their relative performance in multiple limits, as well as numerical results characterizing the conditions under which a crossover in relative advantage appears.
    
    This work determines that for these specific quantum protocols the performance gap between incoherent and coherent access protocol performance is robust to finite noise, although the region in the $\sigma$ (noise strength) and $\delta$ (signal strength) parameterized space of coherent access advantage, as expected, shrinks inverse-polynomially in area with increasing query complexity. We are able to depict this region in terms of these two defining parameters, and verify that numerical approximation of this region agrees with its analytic form in various limits.
    
    Moreover, in analyzing hybrid protocols, we were able to compute, again in suitable limits, the ideal method for distributing probes to the underlying quantum process between coherent and incoherent subroutines for optimal performance in RDGs. For small noise, we find that the optimal coherence length depends on the noise parameter according to the standard quantum limit, while for large noise the dependence becomes more classical in character.
    
    This work represents a new application of QSP methods to noisy settings, and a concrete series of methods for numerically and analytically investigating the robustness of these protocols to reasonable noise. Moreover, this work situates this realistic physical question of robustness within recent work regarding the complexity-theoretic characterization of quantum algorithms in differing access models \cite{qualm_2021}. We also provide a complement to the recent quantum machine learning work in \cite{Huang21}, proposing a separate noise model and classification scheme which is not amenable to their analysis of adaptive protocols, but which is amenable to generalization to large-dimension channels through the methods of QSVT. While the channels considered in this work are simple, they are minimal examples for which QSP-based methods can completely characterize the presence of quantum advantage.
    
    A major caveat of this work, and various works considering the relative performance of specific quantum algorithmic models, is the non-exhaustive characterization of incoherent and coherent access protocols. We do not consider adaptive hybrid protocols, or entangled protocols, which are known in general to offer performance improvements for certain problems \cite{takagi, takagi_general}. However, for the limits we do consider, small noise and small mean separation, the limited protocols we consider are asymptotically optimal, and give good intuition for relative performance.
    
    While there is reasonable evidence bounding the optimal performance of incoherent access adaptive protocols in terms of the performance of their non-adaptive counterparts, these bounds are in general difficult to compute outside specific noise-models (e.g., Haar-random channel access, as in \cite{qualm_2021}), and for more exotic noise models (e.g., non-Markovian channel noise) whose allowed quantum advantage may be more interesting. Consequently, the question of the robustness of the advantage of coherent access protocols over incoherent access ones for general noise models remains open. That said, under additional assumptions or for more structured problems, such as the case of correlated noise or asymmetric hypothesis testing, our finding of a classical character for the scaling of the coherence length in Section~\ref{sec:large_n_limit} need not hold, leaving the door open to more powerful statements of quantum advantage.

\section*{Acknowledgements}

    The authors thank Yoshihisa Yamamoto for useful discussions and valuable comments. 
    This material is based upon work supported by the U.S. Department of Energy, Office of Science, National Quantum Information Science Research Centers, Co-design Center for Quantum Advantage (C2QA) under contract number DE-SC0012704.  We also acknowledge NTT Research for their financial and technical support.
    ZMR was supported in part by the NSF EPIQC program.

\bibliography{main}    

\appendix

\section{Adaptive incoherent strategy}
\label{appendix:adaptive_strategy}

    We consider adaptive incoherent access protocols which based on Bayesian inference as referenced in the main text. Let the prior probability that $\Theta_b$ ($b=0,1$) is the correct hypothesis underlying an RDG be $P(\Theta_b)$ where $P(\Theta_0)+P(\Theta_1)=1$. Moreover, before the first measurement, take $P(\Theta_b) = 0.5$. 
    
    We update this prior every time before next run using Bayes' theorem. To be more concrete, we denote the basis of the adaptive projective measurement by
    $\psi_0$ and $\psi_1$, corresponding to $\Theta_0$ and $\Theta_1$, respectively. We denote conditional probabilities $P(A\mid B)$ in the usual way.
    
    The conditional probability we're interested in is:
        \begin{align}
            P(\Theta_b|\psi_{b'})&=
            \frac{P(\psi_{b'}|\Theta_b) P(\Theta_b)}{P(\psi_b)}, 
        \end{align}
    where 
        \begin{align}
            P(\psi_b)=\sum_{b'=0}^1 P(\psi_{b}|\Theta_{b'})P(\Theta_{b'}).
        \end{align}
    From the prior probability $P(\Theta_b)$ and the conditional probability $P(\psi_{b'}|\Theta_b)$, which we can analytically calculate, we update the probability $P(\Theta_b|\psi_{b'})$ after each measurement. After all measurements have been completed, we threshold to determine the most likely $\Theta_b$. 
    
    In this procedure, there is a degree of freedom for choosing the measurement basis, and a good measurement basis one for which states represented by choice of projector $\psi_b$ are strongly correlated to the hypothesis distributions $\Theta_b$. This correlation is made concrete in the maximization of the mutual information:
        \begin{align}
            I(\Theta;\psi) 
            =\sum_{b,b'} P(\Theta,\psi) 
            \ln\left(
                \frac{P(\Theta,\psi)}{P(\Theta)P(\psi)}
            \right), 
            \label{eq: mutual info} 
        \end{align}
    where $P(\Theta, \psi)$ is the joint probability distribution $P(\Theta, \psi) = P(\psi_{b'}|\Theta_b) P(\Theta_b)$. In Figure \ref{fig:qvh0_theory}, we numerically perform this maximization and plot the resulting discrimination performance. 


\section{Numerical methods}
\label{appendix:numerical_methods}

    The error probabilities that are plotted in Figs \ref{fig:qvh0_theory} and \ref{fig:noisy_case}, are generated by calculating or simulating the probability at each point on an $0.001 \times 0.001$ grid individually. The relevant code is available in the \texttt{noisy-qsp-rdg} repository on Github \footnote{\url{https://github.com/mitquanta/noisy-qsp-rdg}}.
    
    The classical probabilities can be computed via analytical integration. Given two distributions with PDFs $\Theta_0$ and $\Theta_1$, the success probability is given analytically by
        \begin{equation}
            \int_{-\infty}^\infty \frac{1}{2} \left(
        \Theta_0(\theta) \cos^2(\alpha + \theta) +
        \Theta_1(\theta) \sin^2(\alpha + \theta) \right) d\theta,
        \end{equation}
    where $\alpha$ is chosen to maximize the success probability. Here $\cos^2(\alpha+\theta)$ is the probability of successfully measuring $0$ if $\Theta_0$ is chosen and $\sin^2(\alpha+\theta)$ is the probability of successfully measuring $1$ if $\Theta_1$ is chosen. For normal distributions $\Theta_0 = \mathcal{N}(0, \sigma^2)$ and $\Theta_1 = \mathcal{N}(\delta, \sigma^2)$, the integral evaluates to
        \begin{equation}
            \frac{1}{4} \biggr(\left[1 + e^{-2\sigma^2} \cos(2\alpha) \right] +  
            \left[1 - e^{-2\sigma^2} \cos(2(\alpha+\theta)) \right] \biggr),
        \end{equation}
    with the optimal $\alpha = \frac{\pi}{4} - \frac{\delta}{2}$ given by the Helstrom bound \cite{helstrom}. The error probability is equal to the complement, and majority votes are calculated with Eq (\ref{eq:maj-vote}). 
    
    For a given list of QSP-$N$ phase angles $\Phi$ with prepared initial and final states $\psi$ and $\psi'$, the success probability of the QSP protocol $Q_{\Phi}$ is
        \begin{align}
            p_{\Phi, \psi, \psi'} = \int \Theta(\theta_0) \Theta(\theta_1) \Theta(\theta_2) \lvert\braket{\psi' | Q_{\Phi}(\theta) | \psi}\rvert^2 \,d\theta_0\,d\theta_1\,d\theta_2.
            \label{eq: E integral} 
        \end{align}
    However, analytically integrating this is difficult, so we numerically calculate this via Monte Carlo simulations. Specifically, we evaluate the success probability by random sampling of $\theta$: 
        \begin{align}
            p^{\text{approx}}_{\Phi, \psi, \psi'} = 
           \frac{1}{N_r}
            \sum_r 
            \lvert \Braket{\psi' | Q_\Phi(\theta_r) | \psi} \rvert^2 
            \label{eq:eapproxtwo}
        \end{align}
    where $\theta_r$ $(r=1,2,\cdots, N_r)$ are realizations prepared according to the PDF of $\Theta$. Note that $p^{\text{approx}}$ converges to the exact probability $p$ when $N_r \rightarrow \infty$.
    
    Another merit of using \eqref{eq:eapproxtwo} instead of \eqref{eq: E integral} is that we can readily optimize $\Phi$ on a computer, e.g. via \texttt{scipy.optimize}. We optimize both the set of QSP angles $\Phi$ and the prepared states $\psi$ and $\psi'$ to minimize the error probability. Since there are many local minima given $N+4$ degrees of freedom, a single run may not generate the global optimum. Hence we run the optimization several times and take the pointwise best set of angles.

\section{Majority vote for Helstrom bound} \label{appendix:maj_vote_proof}

    We compute the success probability of a fully incoherent non adaptive protocol for an RDG. Given access to a binary process whose success probability follows, in the limit of small $\delta$ the form
        \begin{equation}
            p = \frac{1}{2}(1 + \delta)e^{-2\sigma^2},
        \end{equation}
    then the expected success probability of the classical statistical process which takes $2 M + 1$ samples from this Benoulli distribution and performs majority vote is
        \begin{equation}
            p^\prime = \frac{1}{2}\left(1 + \delta 2^{2M}\frac{(2M + 1)!}{(M!)^2}\right)e^{-2(2M + 1)\sigma^2}.
        \end{equation}
    Proof of this statement follows from applying the majority vote function to $p$
        \begin{equation}
            \sum_{j = 0}^{M + 1} \binom{2M + 1}{j} p^{2M + 1 - j}(1 - p)^j,
        \end{equation}
    and keeping only the first-order terms in $\delta$ (equivalently assuming $\delta$ small). This results in two terms when the sums are collected
        \begin{align}
            2^{-(2M + 1)}&\left[\sum_{j = 0}^{M + 1}\binom{2M + 1}{j}\right] +\\
            2^{-(2M + 1)}&\left[\sum_{j = 0}^{M}\binom{2M + 1}{j}(2M + 1 - 2j)\right],
        \end{align}
    the first of which is simply $1/2$ by the known symmetry and total sum of the binomial coefficients, and the later of which is a known identity in combinatorics providing the partial sum of $j$ against $(2M + 1)C(j)$ (the choose function). It is an interesting generalization to this method to compute the higher order $\delta$ terms, which have similar character.

\end{document}